\documentstyle[aps,preprint]{revtex}
\begin{document}

\draft
\title{Finite temperature many-particle theory of condensed
        \\matter systems in the functional Schr\"odinger picture}

\author{Hyun Sik Noh}
\address{Hanlyo Sanup University, Kwang-yang, Cheonnam 545-800, Korea}
\author{Sang Koo You and Chul Koo Kim\footnote{e-mail: ckkim@phya.yonsei.ac.kr}}
\address{Department of Physics and Institute for Mathematical Sciences, 
         Yonsei University, Seoul 120-749, Korea}
\maketitle

\begin{abstract}

A finite temperature many-particle theory of condensed matter systems
is formulated using the functional Schr\"odinger picture. Using
the interacting electron gas as a model system, we solve the equation
of motion for the density matrix variationally with a Gaussian type
trial density matrix. We show that the present formalism yields
the finite temperature Hartree-Fock
results both for the para- and ferromagnetic states
in a simple and convenient fashion. Implications of
the present results and future prospects are also discussed.

\end{abstract}


\section{Introduction}
The many particle systems at finite temperatures are generally studied
using the temperature Green's function through either diagrammatic or
functional integral approach \cite{one,two,three}. This Green's 
function method
utilizes the Heisenberg picture and enables us to evaluate the
equilibrium and nonequilibrium
thermodynamic properties of systems such as the free energy and the
linear responses
when perturbative expansions are available.
On the other hand, the Schr\"odinger picture, which is a more familiar
concept in quantum mechanics has not been much used in the many-body
theory, although the usefulness of this picture for the study of
quantum structures of field theories was recognized early by several
authors \cite{four,five,six}.
Only recently, it has been shown that a many-particle theory
of condensed matter systems
can be successfully formulated using the functional Schr\"odinger
picture(FSP) \cite{seven}. 
However, the theory was limited to the zero-temperature
and, thus, requires extension to finite temperature to be truly useful
for many-particle studies.

In this paper, we present a finite temperature formulation of many-particle
theory based on the functional Schr\"odinger picture. We apply the theory
on the interacting electron gas system and show that a
Gaussian approximation of the
theory produces the finite temperature Hartree-Fock
results both for the para- and ferromagnetic phases
in a simple and convenient fashion. We also show that the self-consistent
equation for the ferromagnetic splitting yields the Stoner criterion
for the ferromagnetism.

\section{Non-Interacting Electron Gas System}
A finite temperature quantum field theory using the functional Schr\"odinger
picture has been recently formulated \cite{eight,nine}
by generalizing the Gaussian
approximation of the variational calculation in quantum mechanics \cite{ten}.
The method 
is based on the Feynman's observation that the unnormalized
density matrix
\begin{eqnarray}
\rho = e^{-\beta H}
\end{eqnarray}
satisfies the equation
\begin{eqnarray}
-\frac{\partial \rho}{\partial \beta} =H\rho~,
\end{eqnarray}
where $\beta$ is the inverse temperature and $H$ the Hamiltonian of the
system \cite{eleven}.

First, we study the non-interacting particle system which has the grand 
Hamiltonian given by
\begin{eqnarray}
\hat{K} &=&\hat{H}-\mu \hat{N}\nonumber\\
        &=&\sum_{\alpha}\int d^3x \psi_{\alpha}^{\dag}(\vec{x})\bigg[
           -\frac{\hbar^2}{2m}\nabla^{2} \bigg]\psi_{\alpha}(\vec{x})-
           \mu \hat{N}~,
\end{eqnarray}
where $\hat{N}$ represents the number operator and $\mu$ the chemical 
potential.
In the FSP, the fermion field operator $\psi$ and its conjugate momentum 
$i\psi^{\dag}$ are expressed as functional operators.
Following the Floreanini-Jackiw prescription \cite{twelve},
the fermionic field 
operators are expressed as
\begin{eqnarray}
\psi (x) &=& \frac{1}{\sqrt{2}} \bigg[u(x)+
             \frac{\delta}{\delta u(x)^{\dag}}\bigg]~,\nonumber\\
\psi^{\dag} (x) &=& \frac{1}{\sqrt{2}} \bigg[u(x)^{\dag}+
             \frac{\delta}{\delta u(x)}\bigg]
\end{eqnarray}
to satisfy the equal-time anticommutation relation
$\{\psi_i (x,t),~\psi_j^{\dag}(x',t)\}=\delta_{ij}\delta (x-x')$, where
$u$ and $u^{\dag}$ are anticommuting Grassmann variables.
Using the above expressions, we rewrite the grand Hamiltonian as follows,
\begin{eqnarray}
\hat{K} = \frac{1}{2}\sum_{A,B}\xi_{AB}\bigg(u_A^{\dag}+\frac{\delta}
           {\delta u_A}\bigg)\bigg(u_B+\frac{\delta}
           {\delta u_B^{\dag}}\bigg)~,
\end{eqnarray}
where $\xi_{AB}=-\frac{\hbar^2}{2m}[\delta(\vec{x}-\vec{y})\nabla_y^2]
       \delta_{\alpha\beta}-\mu\delta_{AB}$ and $A=(\vec{x},\alpha),~
       B=(\vec{y},\beta)$.
We choose a Gaussian trial density matrix
\begin{eqnarray}
\rho = e^{u_A^{\dag}F_{AB}u_B+J_A u_A+u_A^{\dag}L_A+C}~,
\end{eqnarray}
to satisfy the differential equation,
\begin{eqnarray}
-\frac{\partial \rho}{\partial \beta} =K\rho~.
\end{eqnarray}
Substituting Eq.(6) into Eq.(7), we obtain the following
identity equation,
\begin{eqnarray}
-u^{\dag}\frac{\partial F}{\partial \beta}u-
  \frac{\partial J}{\partial \beta}u
  -u^{\dag}\frac{\partial L}{\partial \beta}
  -\frac{\partial C}{\partial \beta}&=&
  \frac{1}{2}\bigg[ u^{\dag}(\xi-F\xi F)u+u^{\dag}(\xi L-F\xi L)\nonumber\\
     &&+(-J\xi-J\xi F)u+(\xi+F\xi-J\xi L)\bigg]~.
\end{eqnarray}
Since the above equation is an identity equation with respect to 
$u$ and $u^{\dag}$, we obtain four differential equations.
The four equations are readily solved to yield
\begin{eqnarray}
F=-coth(\frac{1}{2}\beta\xi)~,
\end{eqnarray}
\begin{eqnarray}
J=-v^{\dag}e^{\frac{1}{2}\beta\xi}[sinh(\frac{1}{2}\beta\xi)]^{-1}~,
\end{eqnarray}
\begin{eqnarray}
L=-ve^{-\frac{1}{2}\beta\xi}[sinh(\frac{1}{2}\beta\xi)]^{-1}~,
\end{eqnarray}
\begin{eqnarray}
C=Tr\bigg[-\frac{1}{2}\beta\xi+ln[sinh(\frac{1}{2}\beta\xi)]
      -v^{\dag}vcoth(\frac{1}{2}\beta\xi)\bigg]~.
\end{eqnarray}
In the above, Grassmann variables $v$ and $v^\dagger$ are
introduced as constants of integrations in order to make
$\rho$ a Gaussian type density matrix [8].
The normalized density matrix is obtained using these relations,
\begin{eqnarray}
\rho_{nor}=\frac{1}{\cal N}e^{Tr[-\frac{1}{2}\beta\xi]}Det[sinh(\frac{1}{2}
           \beta\xi)]e^{u^{\dag}v-v^{\dag}u}
          e^{-(u^{\dag}+v^{\dag})[coth(\frac{1}{2}\beta\xi)](u+v)}~,
\end{eqnarray}
where the normalization constant $\cal N$ is obtained by the relation,
$1=Tr[\rho_{nor}]$,
\begin{equation}
{\cal N}=e^{Tr[-\frac{1}{2}\beta\xi]}Det[4cosh\frac{1}{2}\beta\xi]~.
\end{equation}
This normalized density operator corresponding to the above normalized
density matrix is given by
\begin{eqnarray}
\hat{\rho}_{nor}=
\frac{e^{Tr[\frac{1}{2}\beta\xi]}}{Det[4cosh(\frac{1}{2}\beta\xi)]}
                 e^{-\beta \hat{K}}~.
\end{eqnarray}
The Helmholtz free energy is defined by
\begin{eqnarray}
\beta F \equiv <ln \hat{\rho}_{nor}>+\beta<H>~.
\end{eqnarray}
Substituting Eq.(15) into Eq.(16) and using Eq.(13) to calculate
the expectation value, we obtain
\begin{eqnarray}
F=N\mu-k_B T~ Tr[ln(1+e^{-\beta\xi})]-Tr[ln2]~.
\end{eqnarray}
The Fourier transformation of the above equation gives the familar
result as follows,
\begin{eqnarray}
F=N\mu-k_B T~ \sum_{k,\sigma}[ln(1+e^{-\beta\xi_k})]-N ln 2~,
\end{eqnarray}
where $\xi_k\equiv \epsilon_k^0-\mu\equiv \frac{\hbar^2k^2}{2m}-\mu$, and
$\alpha$ is the spin index.
The total energy is also readily obtained
\begin{eqnarray}
E&=&<\hat{H}>\nonumber\\
 &=&\sum_{A,B}\int Du Du^{\dag}h_{AB}\psi_A^{\dag}\psi_B\rho_{uv}|_{u=v}\nonumber\\
 &=&\sum_{k,\sigma}\epsilon_k^0n_k~,
\end{eqnarray}
where $ h_{AB} = \xi_{AB} + \mu $ and
\begin{eqnarray}
n_k=\frac{1}{1+e^{\beta (\epsilon_k^0-\mu)}}~.
\end{eqnarray}
These results on the non-interacting electron system clearly show that
a finite temperature many-particle theory can be successfully formulated
using the density matrix method based on the 
FSP approach.
\section{INTERACTING ELECTRON GAS SYSTEM}
Next, we investigate the interacting electron gas system of which the grand
Hamiltonian is given by
\begin{eqnarray}
\hat{K}=\sum_{A,B}\xi_{AB}\hat{\psi}_A^{\dag}\hat{\psi}_B+
          \frac{1}{2}\sum_{A,B}V_{AB}
      \hat{\psi}_A^{\dag}\hat{\psi}_B^{\dag}\hat{\psi}_B\hat{\psi}_A~,
\end{eqnarray}
where $V_{AB}$ is any appropriate particle-particle interaction potential.
The density matrix for an interacting system is not a Gaussian and
generally 
quite complex.
As a first approximation, we choose a trial Gaussian density 
matrix 
with a parameter function, which will
be chosen from minimization of the thermodynamic potential. 
Thus, the normalized trial density matrix is given by 
\begin{eqnarray}
\rho_{nor}=\frac{1}{Det[4 coth(\frac{1}{2}
           \beta Q)]}e^{u^{\dag}v-v^{\dag}u}
          e^{-(u^{\dag}+v^{\dag})[coth(\frac{1}{2}\beta Q)](u+v)}~,
\end{eqnarray}
where $Q$ is the adjustible parameter matrix.
The normalized density operator is written in the form
\begin{eqnarray}
\hat{\rho}_{nor}
=\frac{e^{Tr[\frac{1}{2}\beta Q]}}{Det[4 cosh(\frac{1}{2}\beta Q)]}
                 e^{-\beta \hat{\psi}^{\dag} Q\hat{\psi}}~,
\end{eqnarray}
in analogy with the free particle case.
The thermodynamic potential is defined by
\begin{eqnarray}
\beta {\bf \Omega}&=&\beta(F-\mu N)\nonumber\\
                  &=&<ln\hat{\rho}_{nor}>+\beta<\hat{H}>-\beta\mu<\hat{N}>
                     \nonumber\\
                  &=&<ln\hat{\rho}_{nor}>+\beta<\hat{K}>~.
\end{eqnarray}
Substituting Eq.(23) into Eq.(24), we obtain
\begin{eqnarray}
\beta {\bf \Omega} &=&\frac{1}{2}\beta\sum_A Q_{AA}
                   -\sum_A[ln[4cosh(\frac{1}{2}\beta Q)]]_{AA}\nonumber\\
                   &&-\frac{1}{2}\beta\sum_{A,B}Q_{AB}G_{BA}
                   +\frac{1}{2}\beta\sum_{A,B}\xi_{AB}G_{BA}\nonumber\\
                   &&+\frac{1}{8}\beta\sum_{A,B}V_{AB}[G_{AA}G_{BB}-
                                                       G_{BA}G_{AB}]~,
\end{eqnarray}
where $G=I-tanh(\frac{1}{2}\beta Q)$, and $I$ is the identity matrix.
The Fourier transformation of the above equation is
\begin{eqnarray}
\beta {\bf \Omega} =&-&V\int dK tr[ln[4 cosh A(\vec{k})]]\nonumber\\
                    &+&V\int dK tr[A(\vec{k}) tanh A(\vec{k})]\nonumber\\                   
                    &+&\frac{1}{2}V\beta\int dK tr[\xi(\vec{k})G(\vec{k})]
                        \nonumber\\
                    &+&\frac{1}{8}V\beta V(0)\int dK tr[G(\vec{k})]
                        \int dK' tr[G(\vec{k}')]\nonumber\\
                    &-&\frac{1}{8}V\beta\int dK dK' V(\vec{k}-\vec{k}')
                        tr[G(\vec{k}) G(\vec{k}')]~,
\end{eqnarray}
where $A(\vec{k})=\frac{1}{2}\beta Q (\vec{k})$, 
$G(\vec{k})=\sigma_0-tanh(\frac{1}{2}\beta Q (\vec{k}))$, $tr$ is the
trace for the spin indices only, and $\sigma_0$ is the $2\times 2$ identity
matrix.
Introducing a constraint parameter $N$,
\begin{eqnarray}
N\equiv \frac{1}{2}V\int dK tr[G(\vec{k})]~,
\end{eqnarray}
we rewrite the thermodynamic potential as follows,
\begin{eqnarray}
\beta {\bf \Omega} &=&\alpha\bigg[N-\frac{1}{2}V\int dK tr[G(\vec{k})]\bigg]
                     \nonumber\\
                     &&-V\int dK tr[ln[4 cosh A(\vec{k})]]\nonumber\\
                    &&+V\int dK tr[A(\vec{k}) tanh A(\vec{k})]\nonumber\\                   
                    &&+\frac{1}{2}V\beta\int dK tr[\xi(\vec{k})G(\vec{k})]
                        \nonumber\\
                    &&+\frac{1}{2}\frac{V(0)}{V}\beta N^2\nonumber\\
                    &&-\frac{1}{8}V\beta\int dK dK' V(\vec{k}-\vec{k}')
                        tr[G(\vec{k}) G(\vec{k}')]~,
\end{eqnarray}
where $\alpha$ is the Lagrange's undetermined multiplier.
Taking variations on 
$\beta {\bf \Omega}$ with respect to $N$ and $A(\vec{k})$, we obtain 
\begin{eqnarray}
\alpha =-\frac{V(0)}{V}\beta N~,
\end{eqnarray}
\begin{eqnarray}
\beta Q(\vec{k})=(\beta\xi(\vec{k})-\alpha )\sigma_0-
     \frac{1}{2}\beta\int dK' V(\vec{k}-\vec{k}')G(\vec{k}')~.
\end{eqnarray}
Eq.(30) is a self-consistent equation which determines $Q 
(\vec{k})$.
Since all functions in the above equation are
$2\times 2$ matrices, $Q (\vec{k})$ and $G(\vec{k})$ are also 
represented
as linear combinations of the Pauli matrices.

First, we consider the symmetry conserving paramagnetic case.
In this case, the eigenvalues of $G(\vec{k})$ are same for both spins,
and, thus,
$Q (\vec{k})$ and $G(\vec{k})$ must be multiples of the
$2 \times 2$ identity matrix,
\begin{eqnarray}
Q(\vec{k})=q(\vec{k})\sigma_0~,
\end{eqnarray}
\begin{eqnarray}
G(\vec{k})=g(\vec{k})\sigma_0~.
\end{eqnarray}
Defining $q(\vec{k})\equiv \epsilon_k-\mu$, we can rewrite $g(\vec{k})$ as
follows
\begin{eqnarray}
g(\vec{k})&=&1-tanh[\frac{1}{2}\beta (\epsilon_k-\mu)]\nonumber\\
          &=&\frac{2}{1+e^{\beta (\epsilon_k-\mu)}}\nonumber\\
          &=&2 n_k~.
\end{eqnarray}
Here, $n_k$ is the Fermi distribution 
function of
the interacting system.
The energy spectrum of the interacting system is obtained from Eq.(30)
\begin{eqnarray}
\epsilon_k=\epsilon_k^0+(2S+1)V(0)n-\int dK' V(\vec{k}-\vec{k}')n_{k'}~,
\end{eqnarray}
where $n=N/V$ is the number density. The above equation is the familiar 
self-consistent Hartree-Fock equation
which determines the energy spectrum of the interacting electron gas 
system \cite{one}.   
Using the above results, we readily obtain the total energy and 
the Helmholtz free energy 
\begin{eqnarray}
E=<H>=(2S+1)V\int dK (\epsilon_k^0+\frac{1}{2}\Sigma_k) n_k~,
\end{eqnarray}
and
\begin{eqnarray}
F&=&{\bf \Omega}+N\mu\nonumber\\
 &=&N\mu-Nk_BT ln 2-k_BT(2S+1)V\int dK ln[1+e^{-(\epsilon_k-\mu)/k_BT}]~,
\end{eqnarray}
where the self-energy $\Sigma_k$ is given by
\begin{eqnarray}
\Sigma_k =(2S+1)V(0)n-\int dK' V(\vec{k}-\vec{k}')n_{k'}~.
\end{eqnarray}

Next, we consider the magnetic symmetry breaking solution.
In order to represent the ferromagnetic state, we express $Q(\vec{k})$ 
as follows
\begin{eqnarray}
Q(\vec{k})=(\epsilon_k-\mu )\sigma_0+\gamma_k \sigma_3~.
\end{eqnarray}
Assuming that $\gamma_k$ is much smaller than $\epsilon_k$ and 
expanding to the first order of $\gamma_k$, we obtain
\begin{eqnarray}
tanh \frac{1}{2}\beta\bigg[ (\epsilon_k-\mu) \sigma_0+\gamma_k \sigma_3\bigg]
\approx\sigma_0 tanh\frac{1}{2}\beta(\epsilon_k-\mu)+
       \sigma_3\frac{1}{2}\beta \gamma_k sech^2\frac{1}{2}\beta(\epsilon_k
      -\mu)~.
\end{eqnarray}
Using Eq.(30), Eq.(38) and (39), we obtain for 
the ferromagnetic state
\begin{eqnarray}
\beta[ (\epsilon_k-\mu) \sigma_0+\gamma_k \sigma_3]&=&
(\beta \xi_k-\alpha)\sigma_0-\frac{1}{2}\int dK' 
V(\vec{k}-\vec{k}')\bigg[\sigma_0(1-tanh\frac{1}{2}
\beta(\epsilon_{k'}-\mu))\nonumber\\&&
-\sigma_3\frac{1}{2}\beta\gamma_{k'}sech^2\frac{1}{2}
\beta(\epsilon_{k'}-\mu)\bigg]~.
\end{eqnarray}
From this identity equation, we obtain two self-consistent equations as
follows,
\begin{eqnarray}
\epsilon_{k\sigma}=\epsilon_0+(2S+1)V(0)n-\int dK' 
V(\vec{k}-\vec{k}')n_{k'\sigma}~,
\end{eqnarray}
\begin{eqnarray}
\gamma_k=\frac{1}{4}\int dK' V(\vec{k}-\vec{k}')\gamma_{k'}
sech^2[\frac{1}{2}\beta (\epsilon_{k'}-\mu)]~.
\end{eqnarray}
Eq.(41) shows that the one-particle energy is now dependent on the spin 
index. Solving Eq.(42)
self-consistently, one can obtain the amount of splitting between the up- and
down-energy spectrum. Also Eq.(42) yields the celebrated Stoner condition
as follows
\begin{eqnarray}
1&=&\overline{V}(0)\int dK \left[-\frac{\partial n(\epsilon_k)
 }{\partial \epsilon_k}\right] \nonumber \\
 &\equiv&\overline{V}F(0) \nonumber \\
 &\approx& \overline{V}N(0)~.
\end{eqnarray}
Here, $F(0)$, $N(0)$ and $\overline{V}(0)$ represent the static Lindhard
function, the density of state at the Fermi surface and the average exchange
interaction respectively.
The total energy and the thermodynamic potential are given by
\begin{eqnarray}
E=V\sum_{\sigma}\int dK 
[\epsilon_k^0+\frac{1}{2}\Sigma_{k\sigma}]n_{k\sigma}~,
\end{eqnarray}
\begin{eqnarray}
{\bf \Omega}=-\frac{1}{\beta}ln~2-\frac{1}{\beta}V\sum_{\sigma}\int dK
ln\bigg[1+e^{-\beta(\epsilon_{k\sigma}-\mu)}\bigg]~,
\end{eqnarray}
where
\begin{eqnarray}
\Sigma_{k\sigma}&=&\frac{1}{2}V(0)n-\frac{1}{2}\int 
dK'V(\vec{k}-\vec{k}')n_{k'\sigma}~,\nonumber\\
n_{k\sigma}&=&\frac{1}{1+e^{\beta(\epsilon_{k\sigma}-\mu)}}~.
\end{eqnarray}
$\epsilon_{k\sigma}$ and $n_{k\sigma}$ can be determined self-consistently 
once $\gamma_k$ is obtained through Eq.(42). 

\section{Conclusion}
In this paper, we have presented a finite temperature many-particle theory
based on the functional Schr\"odinger picture formalism.
First, we have shown how the functional Schr\"odinger picture formalism
can be applied to many-particle systems by solving the free-particle system
exactly. Next, using the Gaussian density matrix 
obtained from the 
free-particle system, we have constructed a variational calculation 
formalism for the interacting electron gas system.
The present variational formalism yields both the paramagnetic
and the ferromagnetic solutions as in the case of the zero-temperature 
formalism. For the ferromagnetic state, we have obtained a 
self-consistent equation which determines the splitting of energy between 
the two spin-split states.
Also this equation leads to the Stoner criterion for the ferromagnetism.

Here, we note that the above theoretical scheme is general and, thus,
can be applied to any interacting condensed matter systems.
Since the present theory is variational, and not perturbative, it is
expected to become a useful tool for investigation of strongly correlated
systems. In order to obtain results beyond the Hartree-Fock terms, it is
necessary to carry out calculations beyond the Gaussian approximations.
Variational methods employing non-Gaussian trial functionals and
perturbational approach based on variational basis are under study and
will be reported elsewhere. In conclusion,
we believe that this application of the functional Schr\"odinger picture
theory to finite temperature clearly demonstrates the versality and
possible applicability of the theory to wide range of condensed matter problems.

\acknowledgements
We thank Professors A. Kerman and J. H. Yee and Dr. H. Lee for helpful 
discussions. This work was supported in part by the Korea Ministry of
Education (BSRI-96-2425), the Korea Science and Engineering Foundation
through Project
95-0701-04-01-3 and also through the SRC Program of
SNU-CTP. Also, H.S.N. acknowledges
the Korea Science and Engineering Foundation for support
through the overseas postdoctoral program.

\end{document}